\documentclass[10pt, conference]{IEEEtran}
\usepackage{lineno}
\usepackage[draft]{hyperref}
\usepackage{cite}
\usepackage{amsmath}
\usepackage{graphicx}
\usepackage{epstopdf}
\usepackage{enumerate}
\usepackage{filecontents}
\usepackage{epsfig}
\usepackage{float}
\usepackage{caption}
\usepackage{subcaption}
\usepackage{amssymb}
\usepackage{algcompatible}
\usepackage[linesnumbered,ruled]{algorithm2e}
\IEEEoverridecommandlockouts

\usepackage[usenames,dvipsnames]{color}

\begin{document}

\title{A Survey of Network Requirements for Enabling Effective Cyber Deception}

\author{\IEEEauthorblockN{Md Abu Sayed, Moqsadur Rahman,  Mohammad Ariful Islam Khan,  Deepak Tosh}
\IEEEauthorblockA{Department of Computer Science, University of Texas at El Paso, TX, USA \\
Email: \{msayed, mrahman20, mkhan16\}@miners.utep.edu, dktosh@utep.edu}}

\maketitle

\begin{abstract}	
In the evolving landscape of cybersecurity, the utilization of cyber deception has gained prominence as a proactive defense strategy against sophisticated attacks. This paper presents a comprehensive survey that investigates the crucial network requirements essential for the successful implementation of effective cyber deception techniques. With a focus on diverse network architectures and topologies, we delve into the intricate relationship between network characteristics and the deployment of deception mechanisms. This survey provides an in-depth analysis of prevailing cyber deception frameworks, highlighting their strengths and limitations in meeting the requirements for optimal efficacy. By synthesizing insights from both theoretical and practical perspectives, we contribute to a comprehensive understanding of the network prerequisites crucial for enabling robust and adaptable cyber deception strategies.
\end{abstract}
%
%

\section{Introduction}
Cybersecurity domain faces a wide variety of challenging problems because of the evolving nature of the threats and the complexity of the decisions. One of the most potent threats in cybersecurity is the Advanced Persistent Threat (APT) attack where attackers execute highly targeted, long-term, stealthy attacks against government, military, and corporate organizations. In many cases, APTs successfully establish a deep and persistent presence in a target network for months or even years without being detected. Another challenge in cybersecurity is securing highly dynamic, diverse mobile networks against short-term intense attacks, as on the Internet of Battlefield Things (IoBT) \cite{horak2019optimizing}. 

Recently, cyberattacks frequency has increased, for example, distributed denial of service (DDoS) attacks, represent more than 20 percent of all attacks. Such attacks cause a lot of damage for companies in terms of resources, revenue, and services. In spite of having plenty of detection and mitigation mechanisms for DDoS, they are not viable because of unable to capture and consider recent shift of attack from attacking a target directly to indirect attack. Additionally, the existing defense mechanisms are focusing on reactive approaches to detect and mitigate attacks, but substantial amount of damage has already been done before taking any initiative by authority \cite{aydeger2016mitigating}.

All systems are built to be secured, but gaining perfect security is unattainable as all systems have vulnerabilities and successful attack can be launched to exploit the vulnerabilities. For example, static configuration parameters and system settings remain unchanged over a long period of time, giving attacker opportunities to infer network settings and launch a successful attack. In order to mitigate such vulnerabilities, moving target defense (MTD) is introducing the notion of change in multiple system dimensions to intensify the uncertainty and complexity of early reconnaissance of attacker. It is a proactive approach that reverses attacker reconnaissance advantage by changing the rules of the game in favor of the defenders \cite{maleki2016markov}.

MTD and Cyber Deception have emerged as two major research fields to address this critical issue. A moving target defense alters some features of a system's configuration on a regular basis, creating uncertainty for the attacker and increasing the cost of reconnaissance. Unlike MTD, cyber deception creates a virtual attack surface for the attacker rather than modifying the system's actual attack surface.

At the same time, cyber deception helps attacker to evade detection which can be in one of the following forms: prevent from a true belief or formulating a false belief. In a stealthy attack where the attacker can behave as a legitimate user and remain undetected. For example, strategic attacker can adapt their behaviors to be undetected by knowing the pre-defined rules of the firewalls or the rule-based intrusion detection system. In terms of formulating false beliefs, attacker launch “sacrificial attacks” to trick the defender by introducing the notion that all viruses have been detected and repelled. Adversarial cyber deception introduces information asymmetry and puts attacker in a suitable position \cite{huang2019dynamic,sayed2022cyber,sayed2023honeypot,sayed2023assessing}.

However, cybersecurity is an uneven battlefield as the unequal status between the attacker and the defender naturally gives the attacker more advantages. First, attacker is successful by knowing and exploiting one zero-day vulnerability whereas defender needs to defend against all attacks to be successful. Second, attack also evolves over time and be sophisticated so that traditional defense mechanism are unable to protect them and need to be upgraded. Third, attacker has plenty of time to study a system while defender knows nothing about attacker until attacker is detected \cite{huang2019dynamic,sayed2022cyber}.

To deal with information asymmetry, defender can reactively deploy intrusion prevention and detection system capable to identify stealthy and deceptive attacks which is way more costly. However, defensive deception provides a cost-effective alternative by introducing deliberate and proactive uncertainties into the system \cite{huang2019dynamic,sayed2023honeypot}.

Software-Defined Network (SDN) offers a flexible and programmable network environment with better control over network as it separates control plane from the data plane.  It enables generalized forwarding in the router compared to previous destination-based forwarding. Generalized forwarding can consider a lot of options from transport, network, and link-layer whereas previous destination-based forwarding is only based on IP address in the network layer. In SDN, network administrator has wider range of access to program data centrally. Different types of control can be applied to SDN switches to configure honeynet \cite{han2016honeymix,islam2019performance}.

Recently, security researchers are giving more focus on developing proactive cyber defense to prevent the well-informed attacker. To carry out proactive cyber defense, they use network elements such as utilize the SDN programmability to facilitate deception in the network, invalidate early reconnaissance, network virtualization, and filter-based approach whereas previously cyber defenses were mostly reactive and unable to take appropriate actions against cyberattack as significant amount of damage has already done.

SDN-based network deception includes MTD to defend against crossfire attacks, prevent fingerprinting, gives fine-grained data control to SDN, anomaly detection to detect salient events \& forward them to monitored virtual machine, deceiving attacker by disrupting network traffic, and network MTD to facilitate Self-adaptive End-point Hopping technique. SDN-based reconnaissance invalidation includes concealing the true configuration of the network to deceive attacker and deceiving adversaries by disrupting network traffic information. SDN offers network virtualization to mitigate insider reconnaissance such as network scanning and prevent DDoS attack on the virtualized wireless network. SDN also provides filtering-based approach which finds out legitimate user queries from spoofing queries.

The resulting adversarial competition and repeated interactions for control of a network between the defender and the attacker can be modeled as a game, where the defender is allocating defensive resources such as honeypots, and the attacker is trying to reach his goal by compromising network resources without being detected. For instance, any security problem can be designed as a game between attacker and defender, consequently solving the game in terms of finding optimal strategies for players actually solves the security problem. Game theory is becoming an increasingly important tool for optimizing cybersecurity resources \cite{horak2019optimizing}.

Machine learning emerges as a pivotal tool that can significantly enhance the deployment and adaptability of deception strategies. By leveraging machine learning algorithms, network defenders can dynamically analyze vast amounts of network data, identifying patterns and anomalies that inform the creation of realistic and context-aware deceptive elements. Machine learning models can predict attacker behavior, enabling the proactive adjustment of deception tactics to counter evolving threats. Additionally, machine learning aids in optimizing the allocation of resources, strategically placing decoys and honeypots to maximize their impact \cite{raju2020predicting,emon2020automatic,zhu2021survey,sayed2017understanding,haque2021covid,emon2020automated,mahmud2023machine,fazle2023novel}.

The most significant component of the deception design is the effectiveness evaluation of cyber deception. The defender makes the aim and purpose of the deception project apparent at the design phase. Furthermore, the project must be compatible with the existing network infrastructure. Then, based on the feedback from the implementation outcomes, the defenders refine the deception project \cite{sugrim2018measuring}.

The rest of this paper is organized as follows. Section \ref{sec:related} discusses the different SDN-based frameworks for cyber deception. Section \ref{sec:open-research} discusses the problem that researchers are still addressing and the challenges they are facing. Finally, conclusions are presented in Section \ref{sec:conclusion}.


\section{Background Research}
\label{sec:related}

In this section, we discuss different techniques for network deception \cite{aydeger2016mitigating,han2016honeymix,neupane2018dolus,ma2016self}. Then, we discuss the modeling of the computer network to invalidate reconnaissance \cite{chiang2016acyds,anjum2021role}. At the same time, we discussed a variety of network virtualization approaches \cite{adebayo2021cyber,achleitner2016cyber} and filtering-based deception \cite{sahri2016protecting}. Finally, we discuss the effective evaluation of different deception frameworks \cite{sugrim2018measuring,wu2020effectiveness,rawat2019performance}.

\subsection{Network Deception}
Aydeger et al. \cite{aydeger2016mitigating} mainly focused on handling crossfire attacks which is a variation of DDoS attack.  Crossfire is an indirect attack where the attacker strives to build the link-map of the network analyzing the traceroute messages, finds the critical links, and then floods the selected links with DDoS attack. This attack can isolate specific area of a network and that area unable to provide any services. 

They incorporate Software-defined networking (SDN) with MTD to alleviate distributed denial of service (DDoS) attacks. Also, some previous work addressed this issue with proactive defense mechanism. But their proposed work is different from existing in two directions 1) detection mechanism to identify source-destination pairs that involved traceroute operation are based on SDN controller; 2) defense mechanism uses detection information to configure switches using SDN to enable potential MTD.

\begin{figure}[htbp]
    \centering
    \includegraphics[width=9cm, height=4cm]{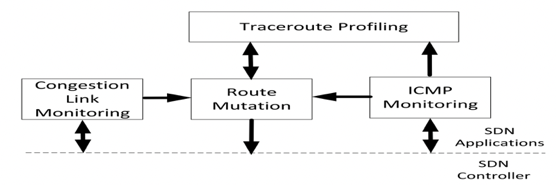}
    \caption{The proposed SDN-based MTD modules \cite{aydeger2016mitigating}}
    \label{fig:crossfire}
\end{figure}

In their defense mechanism, at proactive stage, they obfuscate links during potential link-map creation so that the attacker is unable to lunch the attack and in the reactive stage, they handle detection and mitigation of the attack. From Figure ~\ref{fig:crossfire}, proactive defense consists of ICMP monitoring, traceroute profiling, and route mutation, whereas reactive defense consists of traceroute profiling, route mutation, and congestion-link monitoring. ICMP monitoring tries to detect traceroute operation in the network by observing echo packets and TTL exceeded information in ICMP message. Traceroute profiling application stores detected traceroute info in the database and observes on a particular time any link has excessive traceroute request. Route mutation module takes source, destination, and potential target link as input and tries to find all possible paths from source to destination without that suspected link. Congested link monitoring happened at the SDN controller, if it detects any congested link, then the source-destination pair using that link will trigger route mutation.

Experimental results showed that their proposed SDN- based MTD methodology can effectively handle link flooding by checking and changing routes on a regular basis. At the same time, because of detection and route mutation, MTD will create a bit more delay, especially for those TCP packets.

Honeynet is a collection of honeypots that are used to lure attacker and understand their behaviors and patterns so that can be utilized to thwart the potential attacks. Existing honeypots are suffering from fingerprinting techniques that involve hardcoded timestamps, constant response time, observable features, etc. Another issue with honeynet is coarse-grained data control mechanisms which make honeynet unable to support and communicate with today’s heterogeneous services in honeynet. Conventional architecture only allows one service(honeypot) to interact with the attacker at a time and restricts the defender's ability to lure attacker with multiple services.  For example, Gen-III architecture's existing data control mechanisms are not sufficient to provide that services \cite{han2016honeymix}.

Han et al. \cite{han2016honeymix} overcome the different issues that honeynet faces including fingerprinting and coarse-grained data control. They proposed HoneyMix which utilizes programming functionalities of SDN to attract attacker by simultaneously establishing multiple connections with a group of honeypots. Honeymix is based on traditional Gen-III architecture with honeywall for controlling traffic and SDN for fine-grained data control. Their HoneyMix framework has five core components including Response Scrubber module, Forwarding Decision Engine (FDE), Connection Selection Engine (CSE), Behavior Learner module, and SDN switch. 

\begin{figure}[htbp]
    \centering
    \includegraphics[width=9cm, height=4cm]{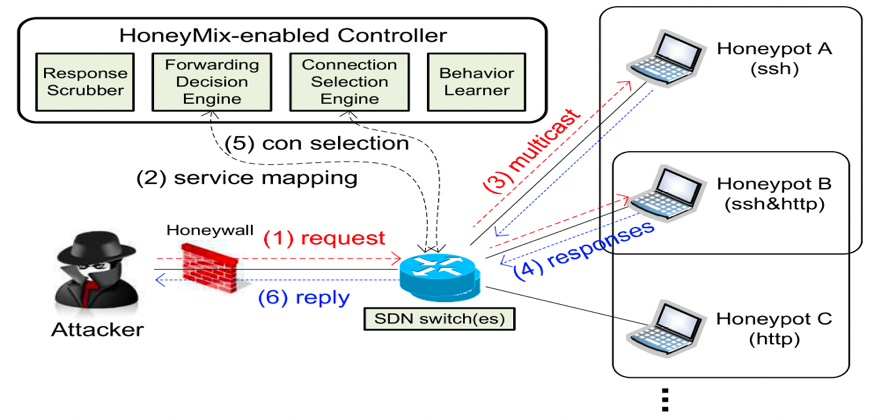} 
    \caption{HoneyMix Architecture \cite{han2016honeymix}}
    \label{fig:honeymix}
\end{figure}

Response Scrubber takes attacker request, especially that involves fingerprinting techniques, and scrubs corresponding responses that reveal honeypots information to attackers. Forwarding Decision Engine (FDE) follows a service map to determine where network service to be forwarded, for example, malicious request has been forwarded to the appropriate honeypot with the help of SDN switches using the Network Function Virtualization (NFV) technique. Connection Selection Engine (CSE) makes the end-to-end connection between the attacker and a honeypot. Behavior computes weights, which CSE uses to select the connection. In weight calculation, connection weight is higher for longer connections and less modification by the response scrubber. SDN switch enables HoneyMix controller to receive data flow and configure network traffic on flight and SDN switch is controlled by FDE. HoneyMix architecture has presented in Figure ~\ref{fig:honeymix}.

Cloud computing has become a critical component of online services available to customers in major consumer industries such as retail, healthcare, manufacturing, and entertainment. These advantages are fueled by advances in cloud platform orchestration that make them completely programmable as Software Defined Everything Infrastructures (SDxI). While SDxI-based cloud adoption is maturing, sophisticated targeted attacks such as Distributed Denial-of-Service (DDoS) attacks are expanding at an unprecedented rate \cite{neupane2018dolus}.

\begin{figure}[htbp]
    \centering
    \includegraphics[width=9cm, height=4cm]{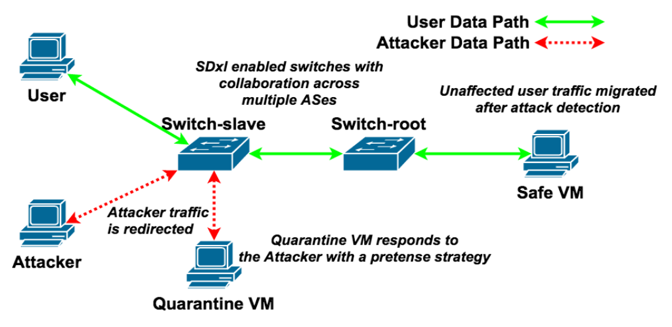} 
    \caption{Illustration of the proposed Dolus system \cite{neupane2018dolus}}
    \label{fig:dolus}
\end{figure}

Neupane et al. \cite{neupane2018dolus} present Dolus as a novel defense system. The Dolus system detects DDoS attacks using threat intelligence derived from attack feature analysis in a two-stage ensemble learning scheme. The first stage focuses on anomaly detection to identify salient events of interest, and the second stage is used to differentiate the DDoS attack event type among the five common attack vectors: DNS (Domain Name System), UDP (User Datagram Protocol) fragmentation, NTP (Network Time Protocol), SYN (short for synchronize), and SSDP (simple service discovery protocol). The Dolus system is unique because it employs a scalable and collaborative defense plan based on foundations from pretense theory in child play, as well as SDxI-based cloud platform capabilities for elastic capacity provisioning via 'quarantine VMs,' and SDxI policy coordination across multiple network domains. A strategy like this is intended to prevent the disruption of cloud-hosted services by deceiving the attacker by creating a false sense of success and preventing the attacker from identifying that a high-value target has been impacted and is being moved. A sample scheme is shown in Figure ~\ref{fig:dolus}. 

Moving target defense (MTD) has been offered as a way to affect the dynamic by removing the attacker's advantage by diminishing the availability of continuous or gradually-changing vulnerability that let attackers to wait and conduct effective tests on persistent flaws. Its goal is to offer a variable, non-sustained and non-deterministic runtime environment. Network MTD (NMTD) uses multi-level adaptive modifications to undermine the attack chains' reliance on the predictable and consistent network environment. End-point hopping is one of the most effective strategies for mitigating network attacks \cite{ma2016self}.

Despite the numerous hopping techniques proposed, existing systems lack the capacity to evolve to various reconnaissance tactics, causing network protection to become blind. Existing end-point hopping research suffers from two major flaws. First, the advantages of hopping defense are reduced due to insufficient network hopping dynamics, which are induced by self-learning inadequacies in the reconnaissance attack approach, resulting in hopping method selection blindness. Second, the availability of hopping mechanisms is inadequate due to restricted network resources and significant overhead.

Ma et al. \cite{ma2016self} propose Network Moving Target Defense based on Self-adaptive End-point Hopping Technique (SEHT), which is based on adversarial strategy awareness and implemented by Software Defined Networking (SDN), to overcome the aforementioned issues. Two features indicate the benefits of this mechanism. By differentiating the scanning attack technique, a hopping trigger based on adversary strategy awareness is provided for directing the choice of hopping mode. To guarantee that hopping has a minimal overhead, satisfiability modulo theories are employed to rigorously specify the requirements. Even in a mixed scanning approach with low-overhead hopping, theoretical analysis and simulation trials reveal that SEHT can withstand roughly 90\% scanning attack.

\subsection{Invalidate Reconnaissance}
Because of various types of vulnerabilities, phishing attacks, and insider attacks, it is now impossible to prevent malicious intrusions using firewalls and other cyber defense mechanisms. Again, increasing the degree of complexity is critical when performing network reconnaissance. However, such activities can be carried out at a very slow pace, and network configurations typically remain static for a long time, making reconnaissance more difficult. As a result, current state-of-the-art intrusion detection systems (IDSs) may be unable to detect zero-day attacks and may generate false alarms \cite{chiang2016acyds}.

\begin{figure}[htbp]
    \centering
    \includegraphics[width=9cm, height=4cm]{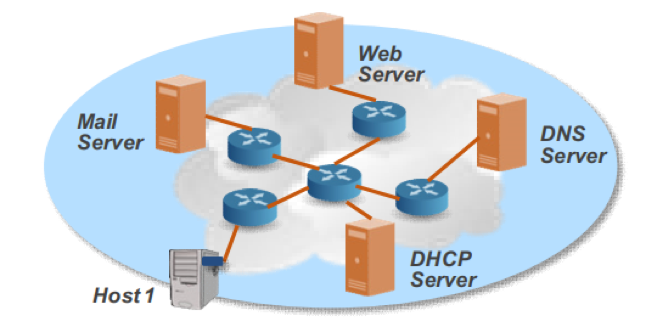}  
    \caption{Host 1’s Network View \cite{chiang2016acyds}}
    \label{fig:acyds}
\end{figure}

Chiang et al. \cite{chiang2016acyds} presented a novel approach called Adaptive Cyber Defense System (ACyDS) generates host a unique virtual network view for each host. Figure ~\ref{fig:acyds} shows a sample. This view includes the IP addresses of reachable hosts, servers as well as subnet topology, etc. But it does not represent the true network configurations. In real-time, it changes the host's network view keeping the difference with other hosts in the network. Because subnet topology and IP address assignments are changed with each view update, ACyDS invalidates intelligence gathered during previous reconnaissance activities. In summary, ACyDS's deception approach prevents collisions if multiple hosts have been compromised, discourages reconnaissance activities, and increases the likelihood and confidence in detecting the presence of intruders. This approach makes use of SDN technologies such as OpenFlow switches and controllers.

Enterprises place a high value on network security, which includes a diverse range of computing resources such as desktops, laptops, servers, routers, and switches. The resources support a wide range of activities carried out by different types of users in different roles (e.g., IT administrators, C-suite executives, and personal). An adversary may cause significant harm to the enterprise by compromising one or more of these resources \cite{anjum2021role}.

\begin{figure}[htbp]
    \centering
    \includegraphics[width=9cm, height=4cm]{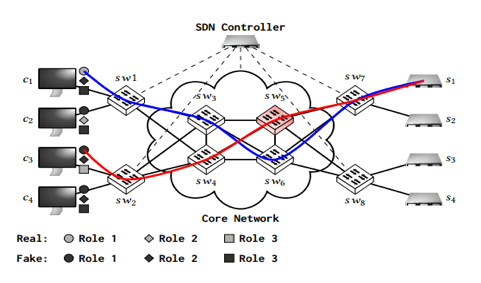}   
    \caption{Overview of HoneyRoles \cite{anjum2021role}}
    \label{fig:honeyroles}
\end{figure}

That’s why Anjum et al. \cite{anjum2021role} focuses on deceiving adversaries by disrupting network traffic information obtained via passive reconnaissance and discouraging an adversary from acting on observed information (e.g., performing active reconnaissance or an attack). The idea is to build metaphorical "haystacks" around these people's network activities. Introduced network traffic disrupts reconnaissance, and if the adversary acts on incorrect intelligence, it will most likely be detected, discouraging the adversary from acting. HoneyRoles, which uses honey connections to deceive adversaries using compromised packet forwarding devices for passive reconnaissance, is proposed here. HoneyRoles coordinates honey connections by simulating fake hosts organized into roles that correspond to the organizational functions of client hosts. HoneyRoles verifies the integrity of honey connections so that they can serve as "canaries" for network client attacks. HoneyRoles detects the presence of an adversary and statistically identifies any compromised forwarding devices when an adversary modifies or blocks a honey connection. Figure ~\ref{fig:honeyroles} shows an overview of HoneyRoles.

\subsection{Network Virtualization}

In deception-based cyber defense systems, honeypots are very popular across the industry and academia. Honeypots lure attacker and give them the vibe of interacting with regular hosts or network nodes. Additionally, honeypot collect information about the attacker actions and learns their motives \cite{adebayo2021cyber}.

Wireless virtualization technology provides abstraction of wireless network infrastructure and radio spectrum resources. But an adversary can make DDoS attack by exploiting certain activities such as subleasing the radio frequency channels from wireless infrastructure providers to  Virtual Network Operators (VNO)s. As a consequence,  VNOs are unable to perform usual services to its legitimate users \cite{adebayo2021cyber}.

\begin{figure}[htbp]
    \centering
    \includegraphics[width=9cm, height=4cm]{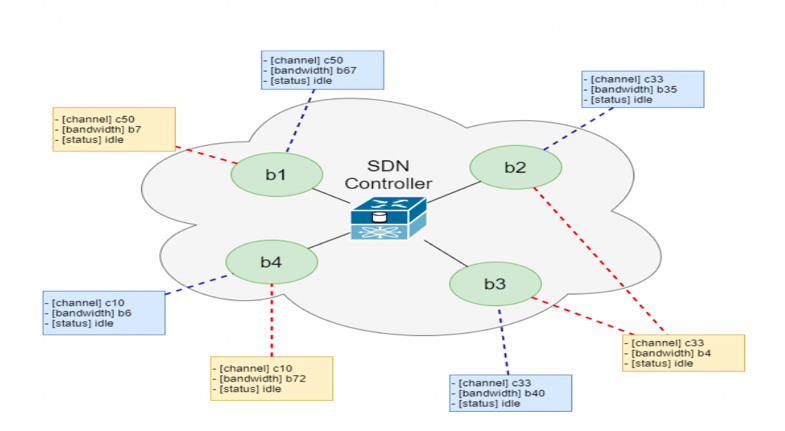}   
    \caption{Sample cyber deception scenario \cite{adebayo2021cyber}}
    \label{fig:wireless}
\end{figure}

Adebayo et al. \cite{adebayo2021cyber} have proposed a cyber deception mechanism in wireless network virtualization for both powerful and naive adversaries. Their naive attacker is not perfectly informed about defender strategies, whereas powerful attacker knows about defender strategies. First, they try to identify the attacker, then move the attacker to a receptor-VNO, and finally, formulate deception using a game-theory approach. In their approach, the defender controls the observable configuration of the network and the attacker takes decision based on this configuration, and they formulate attacker and defender interactions as Stackelberg game. Their proposed cyber defense mechanism showed in Figure ~\ref{fig:wireless}.  

They showed that the computation of an optimal deception strategy is NP-hard for a naive attacker within a budget limit for the cost of deception. They proposed an algorithm for computing an optimal strategy for a naive attacker and a greedy algorithm for selecting the optimal strategy against a powerful attacker. Finally, they showed their proposed algorithms are scalable in terms of low runtime and produce similar defender utilization compared to Mixed Integer Linear Programming.   
 
Network reconnaissance is very useful in static computer networks for attacker to identify potential targets and their vulnerabilities. For example, insider attacker identify hosts, open ports, and map their topology to find known and zero-day vulnerabilities for future attacks. Advanced network scanning is highly effective among network reconnaissance techniques that monitored the uneven distribution of hosts, altering network topologies to accelerate the identification of potential targets \cite{achleitner2016cyber}.

\begin{figure}[htbp]
    \centering
    \includegraphics[width=10cm, height=5cm]{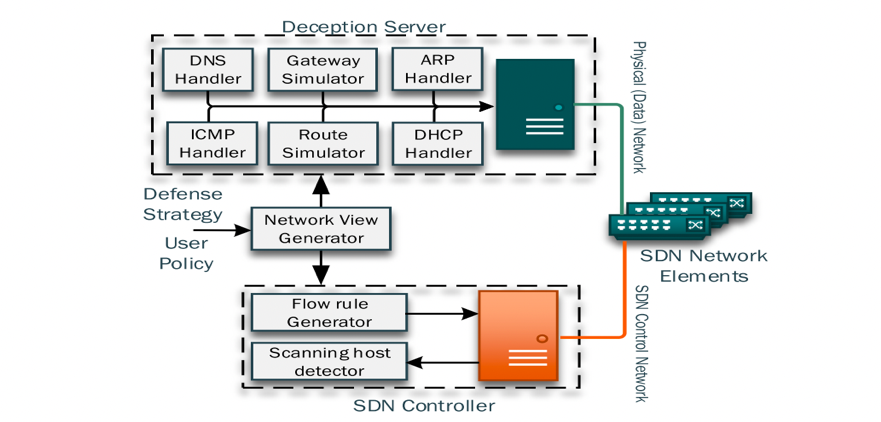}   
    \caption{Architecture of RDS \cite{achleitner2016cyber}}
    \label{fig:rds}
\end{figure}

Achleitner et al. \cite{achleitner2016cyber} formally developed network deception to defend reconnaissance and develop RDS (Reconnaissance Deception System) using SDN controller to achieve deception with the help of virtual network. Their deception server manipulates network traffic and simulates certain virtual network resources considering certain user policies to invalidate attacker collected information in early reconnaissance. Network views generator controls the placement of hosts and honeypots in virtual topologies to delay attackers from identifying real hosts in a virtual topology. Experimentally, they showed their RDS increases the time required to identify vulnerable endpoints in a network up to a factor of 115 with minimizing the performance impact on legitimate traffic on the network to 0.2 milliseconds per packet flow. Finally, SDN dynamically analyzes flow rules and detects scanning activities in the network before scanners find a malicious host in the network. RDS architecture showed in Figure ~\ref{fig:rds}.

\subsection{Token based Filtering}

DNS is primarily a UDP-based protocol that does not require any connection to be established prior to communication between the client and server, making it an ideal tool for hackers to launch distributed denial of service attacks (DDoS). The offered services will not be available to intended users if this method is used. The hackers use a large botnet army to spoof the victim's IP address and make a large number of DNS query attempts in order to flood the DNS server with requests for services. Because the application is designed to accept any IP address range to process the DNS query, the application server cannot tell the difference between an attack and a legitimate query, and as a result, it will simply process all types of DNS queries \cite{sahri2016protecting}.

\begin{figure}[htbp]
    \centering
    \includegraphics[width=9cm, height=4cm]{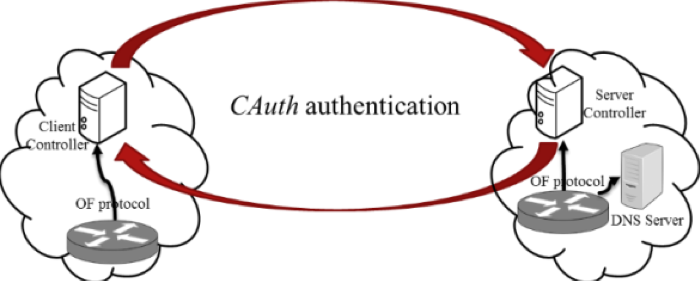}   
    \caption{CAuth Architecture \cite{sahri2016protecting}}
    \label{fig:cauth}
\end{figure}

As a result, to block the “unwanted” DNS query packets, Shari et al. \cite{sahri2016protecting}  proposed a novel mechanism named CAuth. It utilizes Software Defined Networking (SDN), which enables an effective defense against UDP-based DDoS attacks on DNS servers that require detection of source address spoofing. Here the server controller is in charge of sending an authentication packet to each host that has previously requested DNS services. The server controller can determine whether a DNS query launched from any of the source IP addresses is a legitimate query or an attack packet by validating the "authentication packet" that is returned by the requesting client network. The introduced module CAuth can be implemented in DNS application servers without requiring any changes. It can be deployed at any time during DNS server operation without the need for dataset training or manual tuning by administrators. Furthermore, no statistical analysis is used to detect anomalous flow behavior. In its implementation, it makes use of the Openflow protocol's ability to provide a secure communication channel between the network controller and the routers/switches within their network. Figure 2 shows a simplified architecture of CAuth. CAuth is presented in Figure ~\ref{fig:cauth}.

\subsection{Performance Evaluation of Deception System}

The effectiveness of deception defenses has been questioned in a number of studies. Most research, in fact, only looks at single-level qualitative data and overlooks concealability. So, using a dynamic Bayesian attack graph, this research presents a way of evaluating the success of multi-layer cyber deception \cite{wu2020effectiveness}.

In 1988, Pearl created the Bayesian network probabilistic reasoning model, in which nodes are given a starting probability value and directed edges between nodes represent their causal relationship. The attack graph illustrates the possible attack vectors in the network and formalizes the network resources. We can judge the attack path and compute the chance of a node being attacked depending on network resources by using the attack graph based on the Bayesian network \cite{wu2020effectiveness}.

Wu et al. \cite{wu2020effectiveness} assess cyber deception technologies on two levels: concealment of deception initiatives and defense efficacy. In addition, from the attacker's perspective, this study uses the active detection method to compare the consistency between the deception project and the real network assets. The project's concealment is investigated based on the consistency verification results. Then they use the CVSS criteria to assess the attack threat and determine the deception project's effectiveness. They do a simulation experiment to test the practicality of the proposed strategy. Because the honeypots used in this experiment are manually configured, the experiment's deception consistency can be assured. The data show that the overall concealment of the deception project drops slightly as the number of nodes infiltrated by attackers grows. 

Sugrim et al. \cite{sugrim2018measuring} establish the concept of reconnaissance surface as a way to model information that could be revealed to attackers in order to evaluate the success of cyber deception against reconnaissance. Unlike prior studies, they provide metrics to assess the success of network-based deception and use a Bayesian inference model to explain the evolution of an attacker's belief.

They represent an attacker's growing knowledge as a belief system that associates a probability distribution with the existence of each element of the reconnaissance surface during their contact with the target system. The attacker's reconnaissance operations and the defender's deception efforts both influence this belief system. The authors also include suspicion rates in their model to account for different sorts of attackers. The possibility of the information collected during the attacker's reconnaissance is captured by these rates. To see how well network-based deception works against an attacker's reconnaissance efforts, they propose two metrics: 1. the Reconnaissance Surface Measure (RSM), which measures the expected information obtained by an attacker through a series of reconnaissance operations, and 2. the Attacker's Belief Error (ABE), which measures the attacker's final belief's proximity to the ground truth (after a sequence of reconnaissance operations). They consider a network in which a Deception System (DS) is implemented, such as ACyDS. The DS sets a time restriction on the duration of a network view given to a host, restricting the amount of time available to collect data. The DS also actively fools an attacker by deploying honeypots that react to reconnaissance operations for non-existent hosts. The likelihood of fraudulent responses raises the level of uncertainty in the data gathered. The authors demonstrate as a result of the DS's responses to the attacker's queries, weakly held beliefs will be changed quickly, reflecting significant changes in the probability. Strongly held views will take longer to modify. The rate at which an attacker's belief rises relies on the two likelihoods for each response type (positive or negative) (corresponding to the two hypotheses). Normally, these probabilities are evaluated separately. The authors propose that these likelihoods should not be addressed separately in the case of a system that performs a network-based deception technique. 

Since most of the time frequencies are either inactive or underutilized, the static allotment technique of traditional wireless spectrum has been demonstrated to be an ineffective means of utilizing frequentness. This problem has resulted in a artificial RF spectrum shortage. For growing IoT and CPS applications, wireless virtualization is viewed as a way to improve spectrum usage and enable wireless connectivity with high data rates, improved range, and higher quality of experience (QoE). Wireless virtualization allows wireless infrastructure providers (WIPs) and mobile virtual network operators (MVNOs) to tune network parameters on the fly using software defined network (SDN) controllers based on service level agreements (SLAs) to improve end-user QoE. Because of their open nature, cyber-adversaries' malevolent acts are rising dramatically, making it harder to prevent cyber-attacks against wireless networks, which are a primary communication medium for most novel CPS and IoT applications. Virtual wireless networks, as every wireless network, are susceptible to a variety of cyber threats \cite{rawat2019performance}.

Several initiatives in the past in the digital world to prevent cyber attacks employing cyber deception have been prompted by noteworthy characteristics of deception. The primary goal of using deception-based methods is to strengthen conventional cyber defenses by learning from attackers who utilize phony systems and improving the security of real-world systems. Current strategies address cyber deception for traditional network systems. There are no published works that cover the monitoring and assessment of cyber deception in the context of virtual wireless networking. Rawat et al. \cite{rawat2019performance} measure the performance of a cyber deception system in a wireless virtualization framework to counter cyber threats. The SDN-based defender constantly observes network traffic, monitors connections, and constructs deception MVNOs to divert cyber enemies to deception MVNOs and protect authentic MVNOs and their customers from cyber attackers. An Attack Model is considered for this challenge, which comprises of hostile wireless users accessing data wirelessly or overwhelming wireless infrastructure such as MVNO processing and management units. To counteract intruders, the authors propose using a deceptive system to drive cyber attackers to a deception MVNO, ensuring that genuine wireless users are not harmed.

The performance of the suggested strategy is tested using Monte Carlo simulations and numerical findings acquired from formalized mathematical analysis. It can be seen that, for a certain attack arrival rate, earlier deployment of cyber deception not only reduces the number of attacks in MVNOs but also enhances the deception of attacks in virtual wireless networks.

\section{Open Research Challenges}
\label{sec:open-research}

\subsection{Network Deception}

Specific DDoS attack is known as crossfire attack detected by ICMP monitoring and mitigated by route mutation with the help of SDN \cite{aydeger2016mitigating}. Solely depends on traceroute message to identify the congested link to detect crossfire attack is not helpful as from attacker side identification of critical link can be done by following other approaches. Their proactive defense is not fully proactive as does not deploy any techniques such as network deception or virtualization to detect congestion before happening. Also, finding a good threshold for congestion over the network is crucial otherwise it will generate a false alarm when the threshold is very small or cause damage when the threshold is very big. Detection and route mutation create delay on those TCP packets which has an impact on time-sensitive applications. Scalability of detection and route mutation needs more analysis over different test cases and opens up new research directions.

Response scrubber and fine-grained data control modules overhead and their effect on the usual operation of the network need further analysis \cite{han2016honeymix}. Also, a delay sensitivity analysis of the HoneyMix is needed. Evaluation of HoneyMix architecture on real-world deployments will find out any performance-related issues with the architecture.

Dolus uses elastic capacity provisioning in cloud platforms to implement moving target defense techniques that do not affect cloud-hosted application users and contain attack traffic in a quarantine VM (s) \cite{neupane2018dolus}. Although their approach handles DDoS attack in SDN-based cloud infrastructure, adaptation with complex targeted attack is required. For example, advanced data sampling/analysis is required as part of cyber hunting workflows to address more complex targeted attacks such as Advanced Persistent Threats (APTs).

SEHT \cite{ma2016self} hopping has significant overhead, even if it is less than other types of hopping (mutation computational complexity, average transmission delay, and flow table size). It can successfully reduce approximately 29

Token-based filtering which is also known as CAuth can protect against UDP-based DDoS attack \cite{wu2020effectiveness}. It does not require any additional change to the DNS application server, and uses the secure communication channel between the network controller and the routers/switches within their network. The proposed scheme does not necessitate a complex algorithm to detect the spoofed packet. There is no new protocol introduced, and all interactions between the client and server networks are carried out using the standard Openflow protocol, making it a lightweight spoofing detection method. 

But on the other hand, CAuth has opened some challenges. In spoofing detection, it blocks all queries from clients that did not respond to the server controller's previous authentication packet, which might be done selectively. Also, it delays legitimate user request which may affect time-sensitive application, and need further analysis to measure the impact of CAuth on that application in terms of response time. Finally, they work specifically on UDP-based DDoS attack, which opens up new research direction for measuring the applicability of their work on other DDoS attack and extending their work to mitigating other DDoS attacks.

\subsection{Network Virtualization}

Previously published works associate each individual host's view with the port of the switch to which it is connected, rather than with the host itself. The issue of multiple hosts connected to the same port, on the other hand, is not addressed, which has already been addressed in ACyDS\cite{chiang2016acyds}. But generating network view may take a lot of time, which could be improved with a better algorithm, for example - the MAX-SAT solver could be applied here. 

HoneyRoles complements the detection capabilities of prior works by adding a layer of deception and addressing passive (or even subtle active) reconnaissance which was not addressed by the prior work \cite{anjum2021role}. It takes some network request completion time for honeyRoles to reliably rank compromised switches among the most suspicious. This opened up new direction and required further analysis to be attempted to make negligible.

A Reconnaissance Deception System (RDS) is proposed to protect against network reconnaissance \cite{adebayo2021cyber}. As vulnerable hosts are placed far away in the network view, intelligent adversaries can do network scans in reverse order to exploit the vulnerable host. They mentioned in that scenario they will place vulnerable hosts uniformly over the network which is also questionable as how do they know the adversary’s behavior before getting exploit information. Therefore, these issues can be addressed in future research.

\subsection{Performance of Deception System}

Wu et al. \cite{wu2020effectiveness} proposed method is superior. Other works' methodologies all fall under the category of qualitative evaluation, which cannot objectively reflect the success of deception defense. Furthermore, the most significant distinction between this study and others is they often give evaluation methodologies for a specific cyber deception technique, which can be difficult to apply to many deception scenarios. The results demonstrate that the suggested method is capable of presenting attack-defense network events. Apart from that, the consistency verification of the defense project can be used to predict whether the deception would have the desired effect. The defensive efficacy is assessed by extracting and speculating the assault vector in order to aid the defender in performing greater security protection.

However, the strategy used in the paper \cite{wu2020effectiveness} does not take into account specific deception strategies, instead of relying on the findings obtained after implementing the defense project, which may then be applied to a variety of deception scenarios.
Further research should focus on enriching the node architectural model in a complex environment to improve the consistency verification of each node, as well as taking into account the impact of deception defense on typical users when evaluating effectiveness.

Sugrim et al. \cite{sugrim2018measuring} measure the effectiveness of deception system against a variety of attackers and reconnaissance missions.  Their research has important implications: regardless of the type of attacker, there is significant misinformation in the deceptive environment, and the degree of misinformation is strongly proportional to the level of suspicion held by the attackers; highly suspicious attackers are less misguided. Attackers that acquire information with caution increase their yield and remain undetected during an attack. In addition, increased yield entails a wider footprint, which increases the chances of being detected.

However, still there are room for improvement. Different network views with varied degrees of deception and different densities of a property can be examined. The model can also be expanded to include an attacker whose suspicion rates change over time. They used ACyDS to assess the model's efficacy, other analogous deception techniques could be utilized to gain a deeper understanding of the experiment and validate the results in this example.

Rawat et al. \cite{rawat2019performance}  work provides several notable implications such as the fact that all attempts were routed to DMVNO and that no attacks were hindering real MVNOs due to deception. In addition, cyber deception system deployment should be swift enough just to manage attack arrival rates to reduce the attack's impact on the network.

However, they have asserted that MVNOs require some time to identify and report threats to the SDN controller, which is not taken into account in this research. As the time it takes to deploy deception increases, the overall time it takes for attackers to strike MVNOs grows exponentially. If it takes longer to implement deception, attackers will be there for an extended timeframe and attack MVNOs prior to actually tricking them to DMVNO.

\subsection{Advanced Adaptive Attacker}

Advanced adaptive attacker can cause significant damage by collecting and analyzing virtual network views. An advanced adaptive attacker who stays on a host and records a number of virtual network views, can compare these views to determine real and virtual components in network, and estimate the used defense strategy. Additionally, for the coordinated adversary, who exploits multiple hosts, collects their views, consequently, compares them. It is theoretically possible that such attacker can identify the network deception.  With the presence of such an attacker, the generation of virtual network views has to be done with caution, such as alternating different host placement strategies.  Therefore, network view generation as a part of network virtualization with the presence of adaptive coordinated adversaries is an open area of research.

\subsection{Defensive Deception}

Most literature consider defender more intelligent than attacker as only consider simple attack, which is not realistic in practice. Another assumption in Stackelberg game is defender is leader and attacker is follower but in APT attacks such as reconnaissance attack this is not true. So in the future, we need to consider the APT attackers performing multistage attacks.

Existing defensive deception quality is measured by system metrics a proxy measures deployed by defender. But defensive deception is successful when it fully deceives attacker. That defensive deception quality should be based on attacker’s view and actions based on its belief towards the defender’s moves.

In moving target defense, it's really important to keep a balance between security and usability. For example, frequently changing network will give complete security, but which makes network totally useless. That's why in moving target defense, it's important to consider the reconfiguration cost of shifting surface and attacker learning and changing attack vectors. Additionally, moving target defense tries to find an optimal functional configuration of the network for the defender by shifting attack surface that minimizes its risk and damage caused by the attacker.

\subsection{Game Theoretic Analysis}

Also, constructing useful game-theoretic models and applying them in security problems needs to deal with a lot of challenges. Game theory is not applicable in situations where an exact solution is desirable. Constructing a game-theoretic model is difficult because of the complexity of many security domains. Also modeling security problems in the cybersecurity domain is challenging because of the dynamic scenario, and the details of the decision problems are constantly changing. Another issue of the game-theoretic model is it considers all players are perfectly rational, in practice which is not true. That's why behavioral models open new directions for reasoning about deception in security games.

\subsection{Attack Graph Based Analysis}

Attack graph is a graph-based model which is used in cybersecurity research to get simple quantifiable action from complex security scenarios. Attack graph model is also applied to network hardening where a network administrator (the defender) deploys security countermeasures to protect network data from cybercriminals (the attacker).

Defensive deception can be based on an attack graph. Attack graph is a tool used to model network security by capturing network connectivity and vulnerabilities as well as a compact representation of the attacker's plan. Attack graph actually denotes the possible attack path that an attacker can follow to exploit the vulnerability. Interestingly, basic game model can be run on attack graph to provide defensive deception in terms of finding defender strategy.

\subsection{Machine Learning Based Approach}
Machine learning-based approaches help to create accurate fake object and deep reinforcement learning have been considered in developing other defense techniques. Game-theoretic model can model strategic interaction and find approximate solutions. Synergy between machine learning and game-theoretic model can be more effective in developing defensive deceptions. Additionally, machine learning model can be applied to develop MTD. For example, generate viable configuration fake configurations of network using machine learning periodically and replace real configurations with them to detect and prevent attacker.

\section{Conclusion}
\vspace{-0.02in}
\label{sec:conclusion}
In our study, we are trying to find out existing problems that researchers trying to solve in the cybersecurity domain and different approaches they are following to deceive attacker, especially using network components such as SDN controller.

We categorize different deception techniques including network deception, invalidate reconnaissance, network virtualization, token-based filtering, and performance evaluation of deception systems. Network deception is mainly based on MTD which alters the existing network to deceive the adversary. Invalidate reconnaissance actually conceals true network configuration and disrupts network traffic to misinform the attacker. Network virtualization provides different virtual network views to each host to confuse attacker about network to prevent network scanning, DDoS attacks, etc. Token-based filtering is mainly based on authentication tokens to prevent UDP bases DDoS attacks. Performance evaluation of exiting deception framework is focusing on executing different test cases to find viability, scalability, and refine the existing framework.

We also mention some shortcomings of current research \& how they can be improved and a variety of directions through that current SDN-based deception can be extended. Network deception can be improved by considering a good threshold value, analysis delay sensitivity of normal operations, adaptation with the targeted attack, overhead analysis, etc. Network virtualization can be extended through scalable generating of a network view, time sensitivity analysis of request, and generating network view considering the presence of intelligent adversary. Performance analysis of different deception frameworks should consider generalized deception strategies, consistency verification,  analogous deception techniques, response time analysis, etc. Finally, at the time of developing SDN-based cyber deception, some other techniques such as defensive deception, game-theoretic analysis, attack graph-based analysis, machine learning-based approach, etc. should be in consideration.


\bibliographystyle{ieeetr}
\bibliography{References}
\end{document}